\documentclass[%
 reprint,
superscriptaddress,
 amsmath,amssymb,
 aps,
 pra,
]{revtex4-2}
\usepackage{hyperref}
\usepackage{float}
\usepackage{tikz}
\usepackage{lipsum}
\usepackage{braket}
\usepackage{tabularx}
\usepackage{float}
\usepackage{subfigure}
\usepackage{braket}
\usepackage{xcolor}
\usepackage{graphicx}
\usepackage{graphicx}
\usepackage{dcolumn}
\usepackage{bm}

\usepackage{changes}
\definecolor{addedColor}{rgb}{0,0.6,0} 
\definecolor{deletedColor}{rgb}{1,0,0} 
\definecolor{highlightColor}{rgb}{1,1,0} 
\definecolor{commentColor}{rgb}{0,0,1} 

\DeclareMathOperator{\Tr}{Tr}
\usepackage{amsthm}
\newtheorem{definition}{Definition}%
\newtheorem{theorem}{Theorem}

\begin{document}

\preprint{APS/123-QED}

\title{Obtaining Accurate Ground-State Properties on Near-term Quantum Devices}

\footnotetext{\textit{$^{c}$~MOE Key Laboratory of Environmental Theoretical Chemistry, South China Normal University, Guangzhou, 510006, China.}}
\footnotetext{\textit{$^{d}$~Computer Network Information Center, Chinese Academy of Sciences, Beijing, 100190, China. E-mail: yingjin.ma@sccas.cn}}

\author{Qi-Ming Ding}
\affiliation{Center on Frontiers of Computing Studies, Peking University, Beĳing 100871, China}
\affiliation{School of Computer Science, Peking University, Beĳing 100871, China}%

\author{Jiawei Peng}
\affiliation{MOE Key Laboratory of Environmental Theoretical Chemistry, South China Normal University, Guangzhou, 510006, China.}%
\affiliation{Computer Network Information Center, Chinese Academy of Sciences, Beijing, 100190, China.}%

\author{Junxiang Huang}
\affiliation{Center on Frontiers of Computing Studies, Peking University, Beĳing 100871, China}%
\affiliation{School of Computer Science, Peking University, Beĳing 100871, China}%

\author{Yukun Zhang}
\affiliation{Center on Frontiers of Computing Studies, Peking University, Beĳing 100871, China}%
\affiliation{School of Computer Science, Peking University, Beĳing 100871, China}%

\author{Huiyuan Wang}
\affiliation{Center on Frontiers of Computing Studies, Peking University, Beĳing 100871, China}%
\affiliation{School of Computer Science, Peking University, Beĳing 100871, China}%
\affiliation{Peterhouse, Univeristy of Cambridge, Cambridge, CB2 1RD, U.K.}%

\author{Xiaosi Xu}
\affiliation{Graduate School of China Academy of Engineering Physics, Beijing, 100193, China.}%

\author{Jiajun Ren}
\affiliation{Key Laboratory of Theoretical and Computational Photochemistry, Ministry of Education, College of Chemistry, Beijing Normal University, Beijing, 100875, China.}%

\author{Yingjin Ma}
\email{yingjin.ma@sccas.cn}
\affiliation{Computer Network Information Center, Chinese Academy of Sciences, Beijing, 100190, China.}%

\author{Xiao Yuan}
\email{xiaoyuan@pku.edu.cn}
\affiliation{Center on Frontiers of Computing Studies, Peking University, Beĳing 100871, China}%
\affiliation{School of Computer Science, Peking University, Beĳing 100871, China}

\date{\today}

\begin{abstract}
Accurate ground-state calculations on noisy quantum computers are fundamentally limited by restricted \textit{ansatz} expressivity and unavoidable hardware errors. We introduce a hybrid quantum–classical framework that simultaneously addresses these challenges. Our method systematically purifies noisy two-electron reduced density matrices from quantum devices by enforcing $N$-representability conditions through efficient semidefinite programming, guided by a norm-based distance constraint to the experimental data. To implement this constraint, we develop a hardware-efficient calibration protocol based on Clifford circuits. We demonstrate near full configuration-interaction accuracy for ground-state energies of H$_2$, LiH, and H$_4$, and compute precise scattering intensities for C$_6$H$_8$ on noisy hardware. This approach surpasses conventional methods by simultaneously overcoming both \textit{ansatz} limitations and hardware noise, establishing a scalable route to quantum advantage and marking a critical step toward reliable simulations of complex molecular systems on noisy devices.
\end{abstract}
\maketitle

\section{\label{sec:level1} Introduction}
Predicting the behavior of quantum many-body systems is a grand challenge that underpins transformative advances across science and technology~\cite{PhysRevX.14.031006}, from understanding low-energy phases of matter~\cite{PhysRevLett.69.2863} and designing novel catalysts~\cite{pnas} to unraveling the mysteries of high-temperature superconductivity~\cite{RevModPhys.78.17}. Tackling this challenge is stymied by the exponential scaling of the Hilbert space, a ``curse of dimensionality'' that prohibits exact solutions of the Schr\"{o}dinger equation for all but the smallest systems. This has spurred the development of powerful approximation strategies. One path, guided by the Rayleigh-Ritz variational principle~\cite{rayleigh1870finding,macdonald1933successive}, seeks an~\textit{upper} bound to the ground-state energy via a wavefunction \textit{ansatz}. 
The emergence of quantum computing has revitalized this principle, most notably through the Variational Quantum Eigensolver (VQE) and related Variational Quantum Algorithms~\cite{peruzzo2014variational,mcclean2016theory,Cerezo_2021_VQA,TILLY20221}.
Recent theoretical and experimental progress~\cite{kandala2017hardware,kandala2019error,guo2024experimental,ma2025experimental,scienceabb9811} has demonstrated that these methods can enable efficient and accurate determination of ground-state properties for systems beyond the reach of classical computation.

However, the practical implementation of the VQE on current Noisy Intermediate-Scale Quantum (NISQ) devices~\cite{Preskill_2018_NISQ} faces critical obstacles, most notably the restricted depth of quantum circuits and the prevalence of hardware noise. On the one hand, the absence of quantum error correction confines NISQ hardware to shallow circuits, severely limiting the expressive power of quantum ansätze. On the other hand, gate infidelities, decoherence, and measurement errors not only degrade accuracy but can also produce unphysical quantum states.
As a result, pioneering experimental demonstrations have thus far been limited to small molecular systems~\cite{kandala2017hardware,kandala2019error,guo2024experimental}, while extending these methods to larger and chemically realistic problems on practical NISQ hardware remains an outstanding challenge.

Here, we introduce a novel framework that transcends conventional hybrid quantum–classical algorithms~\cite{mcclean2016theory} and error-mitigation techniques~\cite{endo2021hybrid,cai2023quantum,endo2018practical} by simultaneously addressing these two central limitations. By enforcing~\textit{N}-representability constraints~\cite{RevModPhys.35.668,Nakata,PhysRevA.65.062511,cr2000493} through classical postprocessing of raw quantum data, our approach effectively amplifies circuit expressivity while mitigating the noise during the computation and measurement. A norm-based distance between corrected and measured two-electron Reduced Density Matrices (RDMs), calibrated via a hardware-efficient Clifford protocol, provides a systematic and scalable route to robust accuracy.
This framework consistently achieves near Full Configuration-Interaction (FCI) accuracy for molecular ground-state energies, including H$_2$, LiH, and H$_4$, and even surpasses the accuracy of ideal noiseless VQE by overcoming \textit{ansatz} limitations. Extending beyond energetics, we further demonstrate its versatility by reproducing Ultrafast Electron Diffraction (UED) intensities of C$_6$H$_8$ with high fidelity. Together, these results establish a powerful and general strategy for reliable quantum chemistry computation on NISQ devices, charting a scalable path toward practical quantum advantage in chemistry and materials science.

\section{Preliminaries}

\subsection{The Electronic Hamiltonian and Reduced Density Matrices}
The Hamiltonian of a molecule after the Born–Oppenheimer approximation and second quantization can be expressed as
\begin{equation}
H = \sum_{ij} h^{i}_{j}a_i^\dagger a_j + \sum_{pqrs} V^{pq}_{rs}a_p^\dagger a_q^\dagger a_r a_s + H_n,
\label{eq:hamilton}
\end{equation}
where $a^{\dagger}(a)$ denotes the creation (annihilation) operators of the spin orbitals, $h^{i}_{j}$ and $V^{pq}_{rs}$ the one- and two-electron interactions, and $H_n$ collects all non-electron effects such as the interaction between the nuclei. The energy of a state $|\psi\rangle$ with respect to this system is given by the expectation value of the Hamiltonian~(\ref{eq:hamilton}),
\begin{align}
E &= \langle \psi | H | \psi \rangle = \sum_{ij} h^{i}_{j} \langle a_i^\dagger a_j \rangle + \sum_{pqrs} V^{pq}_{rs} \langle a_p^\dagger a_q^\dagger a_r a_s \rangle + H_n \\
&=  \sum_{ij} h^{i}_{j} D_j^i + \sum_{pqrs} V^{pq}_{rs} D_{r s}^{p q} + H_n
\end{align}
where we introduced $\langle \cdot \rangle = \langle \psi | \cdot | \psi \rangle $ as shorthand notation and $D_j^i$ and $D_{r s}^{p q}$ are the 1- and 2-electron RDMs~\cite{RevModPhys.35.668} for a quantum state $|\psi\rangle$, defined as 
\begin{equation}
D_j^i=\langle\psi|a_i^{\dagger} a_j| \psi\rangle,D_{r s}^{p q}=\langle\psi|a_p^{\dagger} a_q^{\dagger} a_s a_r| \psi\rangle
\end{equation}
Measuring these RDM elements to additive error $\epsilon$ generally requires 
$\Omega(N^k/\epsilon^2)$ independent state preparations for a $k$-body RDM on an $N$-orbital system~\cite{Bonet_Monroig_2020_RDM}.  
In practice, one can partition the corresponding Pauli operators into mutually commuting groups, 
so that all terms within each group are estimated simultaneously, thereby reducing the total 
number of measurement settings.

\subsection{The Variational Quantum Eigensolver}

As a hybrid quantum-classical approach, VQE leverages the Rayleigh-Ritz variational principle~\cite{rayleigh1870finding,Ritz1909161,macdonald1933successive} to optimize the lowest possible expectation energy of a trial wavefunction, 
\begin{equation}
    E_0=\min _{\boldsymbol{\theta}}  \langle\psi(\boldsymbol{\theta})|{H}| \psi(\boldsymbol{\theta})\rangle,
\end{equation}
where $| \psi(\boldsymbol{\theta})\rangle$ is prepared on a quantum circuit, serving as an approximation to the ground state energy of a quantum system.
On the one hand, one need a deep quantum circuit to increase the expressiv power of the parametrized state $| \psi(\boldsymbol{\theta})\rangle$. However, the presence of quantum noise means that even with a highly expressive \textit{ansatz}, the experimentally measured 2-RDM, denoted $(D_{rs}^{pq})_{\text{noisy}}$, is noisy and unreliable. Its elements, estimated via Pauli measurements on the quantum device, are not guaranteed to correspond to a physical state and may violate the well-known $N$--representability conditions.

\begin{figure*}[t]
\centering
\includegraphics[width=\textwidth]{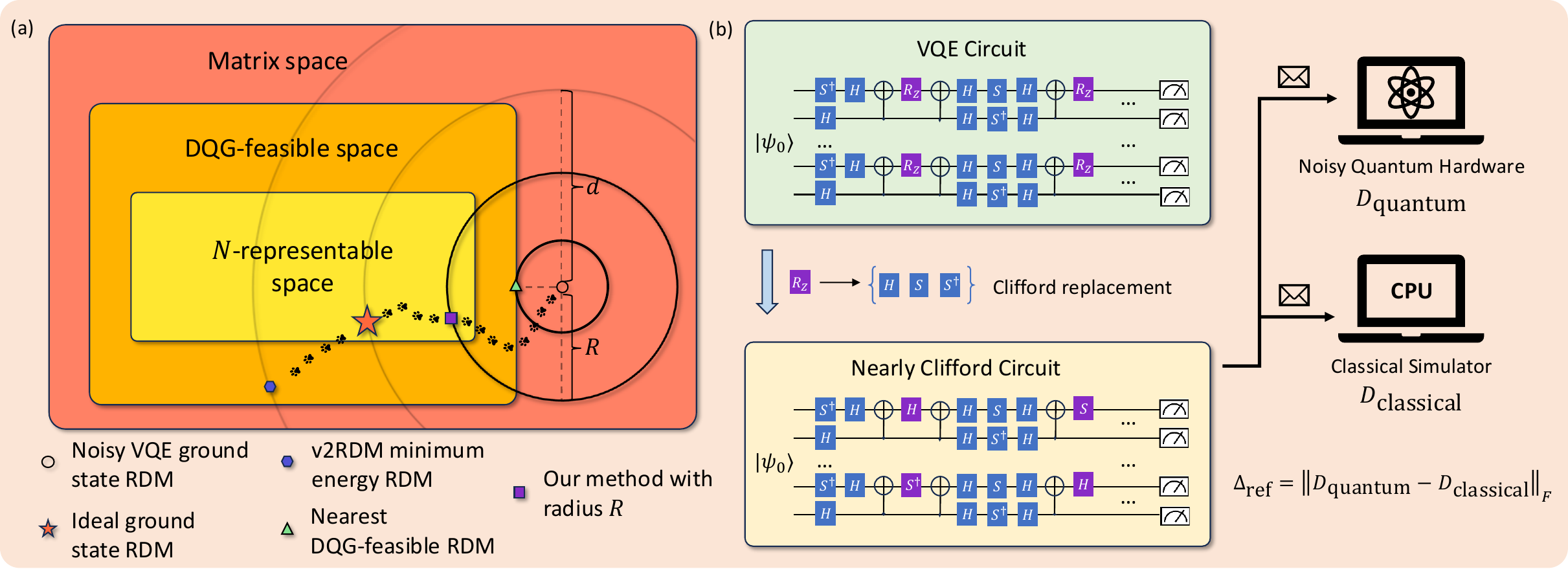}
\caption{
Schematic of the Noise-Aware RDM correction Framework.
(a) A conceptual diagram of RDMs in matrix space. A noisy VQE experiment yields an unphysical RDM ({hollow circle}), which lies at a theoretical (but unknown) distance ${d}$ from the ideal ground state RDM (red star). Our method (purple square) corrects this by finding an RDM that is DQG-feasible (within the orange space) and is constrained to a trust radius $R$ around the noisy result. As shown by the optimization path (paw prints), this process yields a purified RDM that successfully lies within the physical \textit{N}-representable space (yellow). This approach is contrasted with other points like the nearest DQG-feasible RDM (green triangle) and the unconstrained v2RDM global minimum (blue hexagon). {The concentric circles with the hollow circle as the center represent the search results at different search radius $R$.} The correction is most effective when the radius $R$ is comparable to the error distance ${d}$.
(b) The procedure to estimate the unknown error distance ${d}$ and thus determine an appropriate trust radius $R$. A VQE circuit is first transformed into a ``Nearly Clifford Circuit", which is then executed on both the noisy quantum hardware ($D_{\text{quantum}}$) and an ideal classical simulator ($D_{\text{classical}}$). The resulting discrepancy, ${\Delta_{\text{ref}}} = \|D_{\text{quantum}} - D_{\text{classical}}\|$, provides a measurable proxy for the true RDM error. This practical estimate of ${d}$ then guides the selection of the trust radius $R$ for the correction process on the left, ensuring an optimal correction.
}
\label{fig:algorithm}
\end{figure*}

\subsection{Variational 2-RDM Method and N-Representability}
The challenge of unphysical RDMs can be addressed by projecting the noisy data onto the set of valid RDMs. This concept is central to the classical variational 2-RDM (v2RDM) method~\cite{RevModPhys.35.668,Nakata,PhysRevA.65.062511,cr2000493}, which enforces \textit{N}-representability conditions~\cite{Garrod1964,Garrod1975,Erdahl1978,Erdahl1979} within a semidefinite program. Although the v2RDM method provides only a deterministic \textit{lower} bound to the ground-state energy~\cite{PhysRev.97.1474,PhysRev.100.1579,PhysRev.105.1421}, it has proven highly successful across diverse scientific fields, from chemistry and materials science~\cite{Gidofalvi2008, Sinitskiy2010, DePrince2010a, DePrince2010b}, quantum phase transitions~\cite{Gidofalvi2006, Schwerdtfeger2009}, electron-nuclei coupling~\cite{Kamarchik2007pra, Kamarchik2009, Kamarchik2007prl,PhysRevA.65.062511_V2DM,PhysRevLett.106.083001_relatedwork}, and molecular conductance~\cite{Subotnik2009, Rothman2010} to high-temperature superconductivity~\cite{Phillips2010}. More recently, it has also emerged as a powerful tool in quantum computing~\cite{Rubin_2018_NJP_relatedwork,Lanssens_2018_relatedwork,scienceabb9811,piskor2023post,PhysRevLett.130.153001,PhysRevLett.132.220802,anselmetti2024classical}.

The \textit{N}-representability conditions ensure that an RDM corresponds to a valid physical \textit{N}-particle state. A computationally tractable and widely used subset of these conditions is the DQG conditions, which impose positive semidefiniteness constraints on the two-electron ($D$), two-hole ($Q$), and particle-hole ($G$) matrices. The elements of these matrices are defined as:
\begin{subequations}
\begin{align}
    D_{rs}^{pq}&=\langle \psi|a_p^{\dagger} a_q^{\dagger} a_r a_s| \psi\rangle, \label{eq:d} \\
    Q_{pq}^{rs}&=\langle \psi|a_p a_q a_r^{\dagger} a_s^{\dagger}| \psi\rangle, \label{eq:q} \\
    G_{qs}^{pr}&=\langle \psi|a_p^{\dagger} a_q a_r^{\dagger} a_s| \psi\rangle ,\label{eq:g} 
\end{align}
\end{subequations}

Enforcing these conditions, along with other constraints like Hermiticity, antisymmetry, and trace normalization, defines the set of DQG-feasible RDMs. The full set of constraints is as follows:
\begin{subequations}
\begin{align}
    D_{rs}^{pq} &= (D_{pq}^{rs})^* \label{eq:hermitian} \\
    D_{rs}^{pq} &= -D_{sr}^{pq} = -D_{rs}^{qp} = D_{sr}^{qp} \label{eq:antisymmetry} \\
    \sum_{i} D^{i}_{i} &= n ,
    \sum_{pq} {}^2D^{pq}_{pq} = n(n - 1) \label{eq:trace} \\
    D^{i}_{j} &= \frac{1}{N - 1} \sum_{k} {}^2D^{ik}_{jk} \label{eq:contractibility} \\
    D &\succeq 0, \quad Q \succeq 0, \quad G \succeq 0. \label{eq:psd}
\end{align}
\end{subequations}
It is crucial to note that using such a subset is a practical necessity, as the full $N$-representability problem is known to be QMA-complete~\cite{PhysRevLett.98.110503}. This implies it cannot be fully characterized by a polynomial number of constraints (under the assumption of QMA$\neq$P). While stronger approximations like the T1 and T2 conditions exist~\cite{Erdahl1978,Zhao2024}, the DQG conditions offer a robust and computationally tractable approach. Indeed, enforcing them scales polynomially with the number of orbitals $r$, requiring $O(r^4)$ memory and $O(r^6)$ floating-point operations. 

\section{Framework}
\subsection{Noise-Aware Correction of RDMs from VQE}
To address the issue of noisy RDMs from VQE experiments, we introduce a noise-aware RDM correction framework, illustrated in Fig.~\ref{fig:algorithm}. This method employs classical post-processing to purify the noisy quantum data. Our work fundamentally departs from prior RDM correction schemes that merely project a noisy state onto the set of \textit{N}-representable matrices~\cite{Rubin_2018_NJP_relatedwork,scienceabb9811,PhysRevA.100.022517}. Instead of only ensuring physicality, we introduce a controlled search within a defined vicinity of the experimental data to actively seek a more accurate energy. This leads to the following constrained optimization problem:
\begin{equation}
\label{eq:SDP}
\begin{aligned}
& \underset{D_{rs}^{pq}}{\text{minimize}}
& & \text{Tr} \left( K_{rs}^{pq} D_{rs}^{pq} \right), \\
& \text{subject to}
& & \text{\textit{N}-representability conditions (Eqs.~\ref{eq:hermitian}-\ref{eq:psd})}, \\
& & & \left\|D_{rs}^{pq} - (D_{rs}^{pq})_{\text{noisy}} \right\|_{F} \leq \Delta.
\end{aligned}
\end{equation}
Here, the objective function is the energy, where $K_{rs}^{pq}$ is the second-order reduced Hamiltonian~\cite{Nakata},
\begin{equation}
K_{rs}^{pq} = \frac{1}{N-1}\left(h_{r}^{p} \delta_{s}^{q} + h_{s}^{p} \delta_{r}^{q}\right) + V_{rs}^{pq},
\end{equation}
and $\| \cdot \|_{F}$ denotes the Frobenius norm. The crucial innovation is the distance constraint, which defines a hardware-informed trust region of radius $\Delta$, preventing over-correction while respecting theoretical values. For clarity, we refer to this method as the VQE + v2RDM theory.

The role of the Trust Radius $\Delta$ is crucial in this framework. In the limit of $\Delta \to \infty$, the distance constraint is lifted, and our optimization problem reduces to the standard v2RDM calculation, which is guaranteed to yield a lower bound to the true ground-state energy. Conversely, for too small $\Delta$, the trust region may not contain any physically valid RDM, rendering the problem infeasible. A central task, therefore, is to establish a physically meaningful and practically useful value for $\Delta$. To illustrate the critical role of this parameter, we demonstrate in Sec.~\ref{subsec221} how the choice of $\Delta$ impacts the final energy using the LiH molecule as a numerical example.

\subsection{Determining the Trust Radius $\Delta$}

{As a first step toward establishing a physically meaningful value for $\Delta$, one can derive theoretical bounds based on an assumed noise model. We have developed two such bounds, and their formal derivations are presented in the Supporting Information: Methods section.} The practical utility of these theoretical bounds is, however, limited for two primary reasons. First, simplified noise models such as the depolarizing channel fail to capture the full complexity of noise processes in real quantum hardware, which include thermal relaxation, readout errors, and other non-unital coherent effects. Second, as our numerical results will show, the bounds derived from these theorems tend to be overly conservative, yielding a tolerance, $\Delta$, that is too large. An excessively large trust region can cause the optimization to undershoot the true ground-state energy, thereby undermining the goal of our correction framework. These limitations motivate the development of a more practical and data-driven method about quantum hardware parameters to determine an effective $\Delta$.

We introduce a practical, data-driven framework to correct noisy RDMs by enforcing physical consistency within a calibrated tolerance, $\Delta$. Our approach is twofold: first, we benchmark the baseline hardware noise using a reference circuit; second, we use a machine learning model to predict a problem-specific scaling factor that also accounts for \textit{ansatz} approximation errors.

To establish a hardware-specific noise baseline, $\Delta_{\text{ref}}$, for the target VQE ansatz, we implement the following three-stage characterization protocol.

First, we construct a classically simulable reference circuit that mirrors the structure of the VQE ansatz. This is achieved by approximating the ansatz: each non-Clifford single-qubit gate is systematically replaced by its nearest Clifford counterpart. The nearest counterpart is unambiguously determined by identifying the Clifford gate from the candidate set \{H, X, Y, Z, S, S$^\dagger$\} that maximizes the process fidelity with the original gate's operator. This procedure yields a Clifford circuit that is, by design, efficiently simulable classically while preserving the depth and qubit connectivity of the original ansatz.

Next, we characterize two versions of the 2-electron Reduced Density Matrix (2-RDM) for this reference circuit. The ideal RDM, denoted $(D_{rs}^{pq})_{\text{CR\_thm}}$, is computed exactly to high precision via classical simulation. Concurrently, the experimental RDM, $(D_{rs}^{pq})_{\text{CR\_noisy}}$, is obtained by executing the same reference circuit on the quantum hardware and performing the necessary measurements. This provides a direct comparison between the theoretical output and the noisy, hardware-realized output.

Finally, the deviation between the ideal and experimental RDMs provides a quantitative noise baseline that is specific to both the circuit's structure and the hardware's performance. We define this baseline, $\Delta_{\text{ref}}$, as the Frobenius norm of the difference between the two RDMs:
    \begin{equation}
        \Delta_{\text{ref}} = \left\| (D_{rs}^{pq})_{\text{CR\_thm}} - (D_{rs}^{pq})_{\text{CR\_noisy}} \right\|_{F}.
        \label{eq:delta}
    \end{equation}
This metric, $\Delta_{\text{ref}}$, holistically encapsulates the total experimental error—arising from gate infidelities, decoherence, and readout errors—for a circuit that is structurally equivalent to the VQE ansatz.

While $\Delta_{\text{ref}}$ quantifies the hardware noise, it does not account for the intrinsic approximation error of the VQE \textit{ansatz} itself. In this work, we employ the Unitary Coupled-Cluster Singles and Doubles (UCCSD) ansatz~\cite{peruzzo2014variational,PhysRevA.98.022322}, which is known for its high expressivity and chemical accuracy. Because UCCSD provides a robust approximation to the true ground state, its intrinsic error is expected to be small and of a comparable magnitude to the hardware noise captured by $\Delta_{\text{ref}}$. Therefore, to form a comprehensive error bound that incorporates both sources, we introduce a modest scaling factor, $k$, to define the effective tolerance as $\Delta = k \cdot \Delta_{\text{ref}}$. Based on this physical motivation, we set \textbf{$k=2$} throughout our study. This choice establishes a consistent and well-grounded tolerance, positing that the total error is reasonably bounded within a small multiple of the characterized hardware noise.

\section{Numerical Simulation}\label{subsec22}
\subsection{The Role of the Trust Radius $\Delta$}\label{subsec221}
\begin{figure}[ht]
    \centering
    \includegraphics[width=0.45\textwidth]{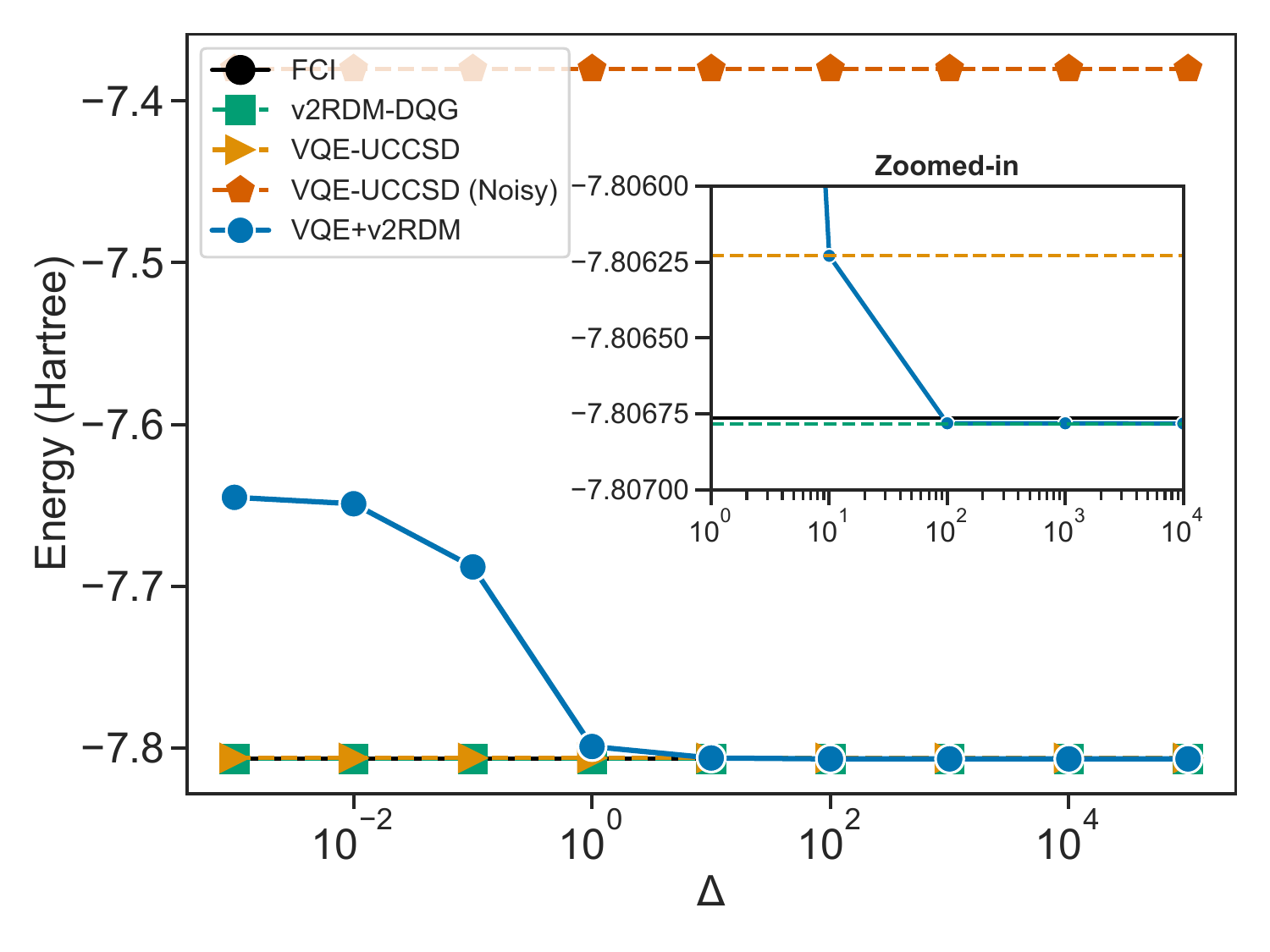}
    \caption{Ground state energy of the LiH molecule at an interatomic distance of $d = 2.8$~\AA. The energy calculated with the our VQE+v2RDM method shown in Eq.~(\ref{eq:SDP}) (solid blue line with markers) is plotted as a function of the penalty coefficient $\Delta$ defined in Eq.~(\ref{eq:SDP}). For comparison, horizontal lines indicate the benchmark FCI energy (solid black), the noisy result of energy from a VQE-UCCSD circuit (dashed orange) {with error rates of \(p_1=0.001\) for single-qubit gates and \(p_2=0.01\) for two-qubit gates}, and the result from the v2RDM method with DQG conditions(dashed green). The energy of VQE+v2RDM theory converges to the FCI value for large $\Delta$. Zoom in view: A magnified view of the converged energy region.}
    \label{fig:SDPexp}
\end{figure}

{We begin by demonstrating the effect of the Trust Radius, $\Delta$, on the calculated ground-state energy of the LiH molecule with the STO-3G basis set at an interatomic distance of 2.8~\AA. To establish clear benchmarks, we first compute the energy using three reference methods: FCI for the exact value, and both noiseless and noisy VQE calculations. The VQE simulations are performed using a UCCSD \textit{ansatz}, employing an active space of three orbitals that are mapped to 6 qubits. For the noisy simulations, we incorporate depolarizing error rates of $p_1 =0.001$ for single-qubit gates and $p_{2}=0.01$ for two-qubit gates. Unless otherwise noted, all numerical examples in this work utilize this specific UCCSD \textit{ansatz} and noise model. With these reference points, we apply our method, as defined in Eq.~(\ref{eq:SDP}), to the noisy VQE results. The outcomes, depicted in Fig.~\ref{fig:SDPexp}, show a clear and systematic trend. As $\Delta$ is increased across a wide range from $10^{-2}$ to $10^{4}$, the resulting energy progressively decreases from the noisy VQE value. The corrected energy is observed to cross the FCI benchmark line and ultimately converges toward the v2RDM lower bound, perfectly illustrating the function of our framework. The benchmark FCI energies, serving as our ``gold standard'', were obtained using the \texttt{fci} module in {PySCF}~\cite{sun2018pyscf,sun2020recent}.}

\subsection{Noise-aware RDM Correction for {Dissociation Energies}}

Modeling molecular dissociation poses a stringent test for any electronic structure method because the nature of electron correlation shifts dramatically along the reaction coordinate. Near equilibrium geometry, the predominant effect is dynamic correlation, which accounts for instantaneous electron repulsion. As the bond stretches towards dissociation, however, static correlation arising from near-degenerate electronic states becomes critical, and the intermediate region exhibits a complex interplay of both.

Faithfully capturing this continuous transformation is a well-known challenge for quantum \textit{ansatz}, whose expressive power is often limited by circuit depth and connectivity on NISQ devices. The results presented in this section demonstrate that our correction framework effectively compensates for these limitations, delivering chemically accurate energies across the entire potential energy curves (PECs). This robustness highlights a key advantage of our approach: it enhances the representation power of the underlying quantum circuit, enabling a unified treatment of diverse correlation regimes that are typically difficult to describe with a fixed, shallow \textit{ansatz}.

{We demonstrate our framework by calculating the dissociation curve for the H\textsubscript{2}, LiH, and linear H\textsubscript{4} molecule, with results benchmarked against exact FCI calculations. For these calculations, we employed the STO-3G basis set. For H\textsubscript{2} and linear H\textsubscript{4}, all orbitals were included, corresponding to simulations on 4 and 8 qubits, respectively. For LiH, an active space of 3 orbitals was selected, resulting in a 6-qubit computation. To validate the versatility of our method, we applied it to two distinct \textit{ansatz} classes: the chemistry-inspired {UCCSD} and a general-purpose HEA. These were specifically chosen to represent two important \textit{ansatz} families: UCCSD is known for its high expressive power derived from chemical principles, while the HEA offers tunable expressivity well-suited for near-term quantum hardware. All VQE simulations were conducted using {Qiskit}'s noisy simulator~\cite{javadi2024quantum}, configured with single-qubit gate error probabilities of $p_1=0.001$ and two-qubit gate error probabilities of $p_2=0.01$. }

\begin{figure*}[ht]
    \centering
    \includegraphics[width=0.95\textwidth]{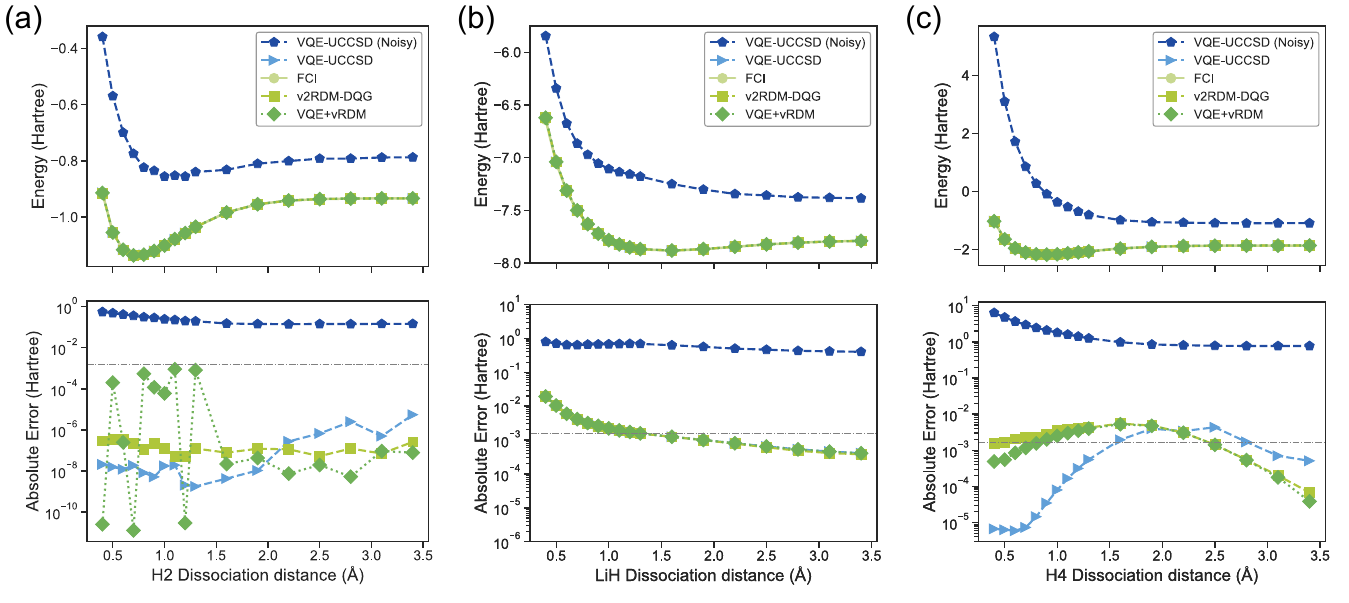}
    \caption{Potential energy curves {and error with FCI} for the dissociation of (a) the H$_2$ (b) the LiH and (c) a linear H$_4$ molecule for UCCSD \textit{ansatz}. In all plots, results from noisy VQE calculations (orange pentagon) are compared against the exact FCI benchmark (solid black line), ideal noiseless VQE simulations (yellow triangle), and our VQE+v2RDM method (blue rhombus). The noisy simulations assume a depolarizing channel with single- and two-qubit gate error probabilities of $p_1=0.001$ and $p_2=0.01$, respectively. The dashed line at $1.6 \times 10^{-3}$~Ha marks the chemical accuracy threshold.}
    \label{fig:H4}
\end{figure*}

We first evaluate our framework using the highly expressive UCCSD \textit{ansatz}. For this demonstration with the H\textsubscript{2} molecule, we used a fixed correction multiplier of $k=2$. The results are presented in Fig.~\ref{fig:H4}. The top panel shows that the raw energies from the noisy VQE simulation ({blue circles}) deviate significantly from the exact FCI benchmark ({light green squares}). In contrast, our corrected energies ({dark green diamonds}) successfully suppress this noise and closely track the true dissociation curve. The bottom panel confirms this, showing that the absolute error for our method remains consistently below the chemical accuracy threshold of $1.6 \times 10^{-3}$~Ha across the most bond dissociation range.

\subsection{Signal denoising in UED simulation }\label{subsec23}

{Having demonstrated the capability of our framework to correct PECs involving bond breaking-processes governed by evolving electron correlation, we now extend its application to another critical area of quantum dynamics: the simulation of experimental observables in UED \cite{centurion2022ultrafast, filippetto2022ultrafast, champenois2023femtosecond}. UED is a powerful time-resolved technique that directly probes atomic-scale structural dynamics during chemical reactions, such as the bond-breaking events illustrated in the previous section. 
However, extracting accurate structural information from UED signals requires highly precise electronic structure inputs, particularly the one- and two-electron RDMs, which are susceptible to noise on quantum hardware. 
}

{For a given wave function $|\psi\rangle$,} the scattering intensity in UED can be simulated using the incoming electron/photon energy $\epsilon_0$ and scattered electron/photon energy $\epsilon_s$ \cite{yang2020simultaneous},
\begin{equation}
   I(\overrightarrow{s},\epsilon_s,i) = \frac{\epsilon_s}{\epsilon_0} \Vert \langle \psi_0 | \widehat{L} |\psi_i \rangle \Vert ^ 2 \delta (E_0 + \epsilon_0 - E_i - \epsilon_s),
\end{equation}
where the $\overrightarrow{s}$ is the momentum transfer vector, $\widehat{L}$ is the scattering operator, $E_0$ and $E_i$ are the energies for the initial ($\psi_0$) and final ($\psi_i$) states, respectively. 
The signal can be expanded into elastic and inelastic components. In the seccond quantization form, these components can be expressed as
\begin{equation}
\label{eq:UED1}
\begin{split}
    I(\overrightarrow{s},\rm{elastic}) = \left |C_{N^{\alpha}} \right |^2 & -2C_{N^{\alpha}}\sum_{ij}D_j^iS_{ij} \\ 
    & + \sum_{ij}\sum_{kl}D_j^iD_l^kS_{i,j}S_{kl}^*,
\end{split}   
\end{equation}
and
\begin{equation}
\label{eq:UED2}
\begin{split}
   I(\overrightarrow{s},\rm{inelastic}) = n & + \sum_{ijkl}D_{kl}^{ij}S_{ij}S_{kl}^*  \\ 
   & - \sum_{ij}\sum_{kl}D_j^iD_l^kS_{i,j}S_{kl}^* ,
\end{split}    
\end{equation}
where the $S$ is diffraction integrals in molecular orbitals basis, the $C_{N^{\alpha}}$ and $n$ are the constants that related to the nucleus and electrons, respectively. 
Intrepreting UED signals requires highly accurate electron correlation models, as the scattering intensity $I(\overrightarrow{s})$ sensitively depends on the many-body electronic structure (Eqs.~\ref{eq:UED1}, \ref{eq:UED2}), particularly the one- and two-electron RDMs, which are susceptible to noise on quantum hardware.

\begin{figure}[htb]
\centering
  \includegraphics[width=0.45\textwidth]{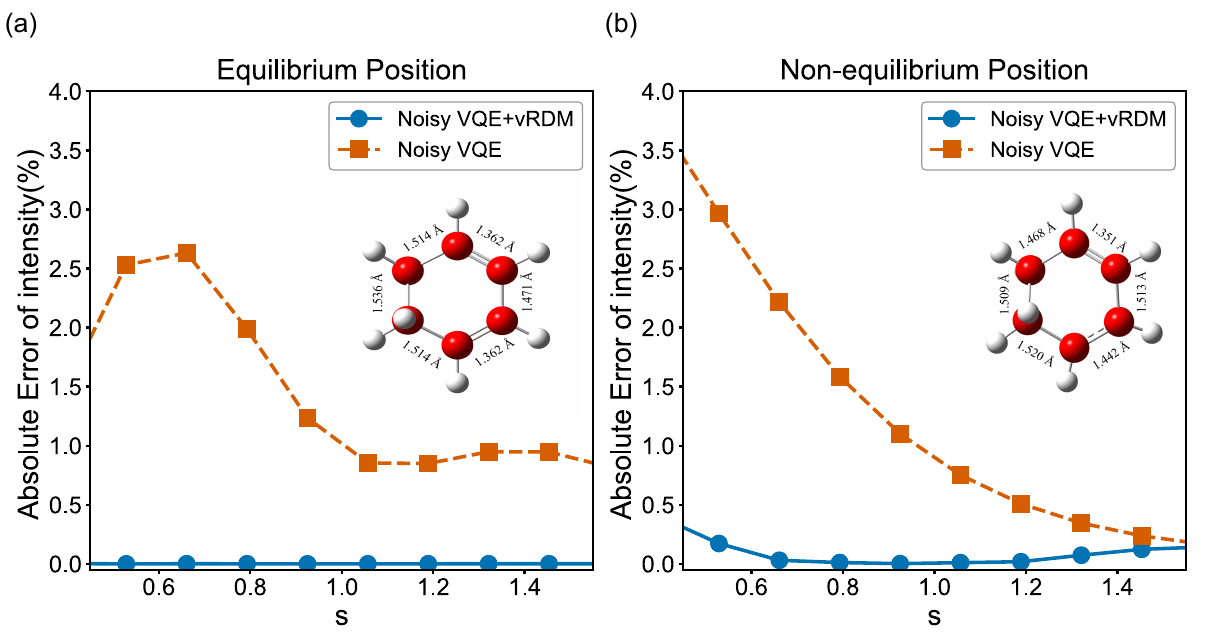}
  \caption{Absolute error of the simulated UED intensity for the C$_6$H$_8$ molecule (a) eauilibrium position (b) non-eauilibrium position. The left panel shows the results at the equilibrium position , and the right panel shows the results at a non-equilibrium position. The orange dashed line with square markers ('Noisy VQE') indicates the error of the spectrum calculated on a noisy quantum computer relative to the exact, noiseless solution. The blue solid line with circle markers ('Noisy VQE+vRDM') represents the error after applying our proposed VQE+vRDM correction method.}
  \label{fig:CHD}
\end{figure}

{By applying the same noise-aware correction protocol, we show that our method significantly denoises UED signals simulated from noisy quantum outputs, thereby establishing a reliable quantum-classical pipeline for predicting cutting-edge experimental observables.}
The UED signal is shown in Fig.~\ref{fig:CHD}, demonstrating that our purified RDMs reduce errors very obviously. Since the signal intensity approaches zero for small $s$, we rescrict our analysis to the range of $s$ = 0.5-1.5 to optimize computational efficiency. 

\section{Conclusion}\label{sec13}
In this work, we have established a quantum-classical framework that systematically refines the noisy outputs of quantum computers to determine ground-state properties with high precision. The method enforces $N$-representability conditions on the experimentally measured 2-RDM within a trust region, which is defined by a norm-based distance constraint. A practical, hardware-efficient protocol for calibrating this trust region was introduced, leveraging classically simulable Clifford circuits to establish a direct link to the hardware's noise characteristics. The efficacy and high fidelity of this approach are validated through numerical simulations on molecules including H$_{2}$, LiH and H$_{4}$, well as by the accurate computation of the UED scattering intensity of C$_6$H$_8$.

The significance of this work extends beyond conventional error mitigation. The presented framework establishes a new paradigm that simultaneously corrects for hardware noise and overcomes the intrinsic expressive limitations of the quantum \textit{ansatz}. By synthesizing the exploratory power of quantum computation with the mathematical rigor of classical constraints, it transforms unreliable data from NISQ devices into physically meaningful and chemically accurate results. This provides a robust and systematic solution to the critical challenge of extracting reliable information from imperfect quantum processors. A key strength of this framework is its generality; although our demonstrations utilized standard DQG $N$-representability conditions and semidefinite programming, the architecture is inherently flexible, designed to seamlessly incorporate more advanced constraints and state-of-the-art algorithms from vRDM theory~\cite{PhysRevLett.117.153001,fosso2016large,PhysRevA.102.052819}. Furthermore, the scalability of the classical post-processing, especially when augmented with symmetry considerations, suggests a viable path toward treating systems of over 30 electrons~\cite{PhysRevLett.117.153001}. We anticipate that this foundational approach will become an essential tool in the ongoing quest to harness the power of near-term quantum computers for scientific discovery.

\section*{Data availability}
The Python code used to perform the simulations and generate the figures in this study, along with the resulting data, have been deposited in the Zenodo repository and are available at https://doi.org/10.5281/zenodo.17277816

\begin{acknowledgments}
This work is supported by the Innovation Program for Quantum Science and Technology (No.2023ZD0300200), the National Natural Science Foundation of China (No.12361161602) and NSAF (No.U2330201). Y.M. acknowledges support by the National Natural Science Foundation of China (No.22173114, 22333003) and the Strategic Priority Research Program (No.XDB0500101), Youth Innovation Promotion Association (No.2022168) of Chinese Academy of Sciences. The scientific calculations in this paper supported by the High-performance Computing Platform of Peking University and ORISE and ERA supercomputers of Chinese Academy of Sciences. 
\end{acknowledgments}

\bibliography{apssamp}

\newpage
\appendix
\begin{widetext}
\section{Supporting Information: Methods}

In this section, we present and prove two theorems that establish theoretical bounds on the Trust Radius, $\Delta$. To formalize this, we first define the error, $\Delta$, as the Frobenius norm of the difference between the ideal RDM from a state $\rho$ and the experimentally measured noisy RDM.
\begin{definition}
    Let $(D_{rs}^{pq})_\text{ideal}$ denote the ideal 2-RDM resulting from a quantum state $\rho$, and let $(D_{rs}^{pq})_\text{noisy}$ represent the 2-RDM obtained from measurements performed on a quantum device. The error $\Delta$ between the ideal and noisy measurements is defined as
    \begin{equation}
        \Delta(\rho)\equiv \sqrt{\sum_{rspq}\left((D_{rs}^{pq})_\text{ideal}-(D_{rs}^{pq})_\text{noisy}\right)^2}.
    \end{equation}
\end{definition}

Our first theorem establishes a bound on this error based on a local depolarizing noise model, where each gate contributes independently to the total error.
\begin{theorem}
    Let a quantum circuit consist of $n$ qubits, with $n_1$ single-qubit gates and $n_2$ two-qubit gates. Assume the presence of local depolarizing noise in the circuit, with error probabilities $p_{k_1}$ for single-qubit gates and $p_{k_2}$ for two-qubit gates. Let $\rho$ denote the quantum state corresponding to the ideal circuit. Then, the error $\Delta(\rho)$ between the ideal and noisy measurements satisfies a bound
    \begin{equation}
        \Delta(\rho) \leq 2n^2\sum_{k_1, k_2}^{n_1, n_2}(p_{k_1}+p_{k_2}).
    \end{equation}
    \label{thm:1}
\end{theorem}

While Theorem~\ref{thm:1} provides a general upper bound, its dependence on the sum of all individual gate errors can be loose. To derive a more structured and potentially tighter bound that depends on the circuit's depth, we now introduce a global noise model.
\begin{definition}
    Define a global depolarizing noise model as follows: after each layer of single-qubit and two-qubit gates, the entire quantum state undergoes the following channel
    \begin{equation}
        \Phi_i\left(\rho\right)=\left(1-p_{i_k}\right) U_i \rho U_i^{\dagger}+p_{i_k} \frac{I_n}{2^n}, 
    \end{equation}
    where $k = 1, 2$ represents the cases for single- and two-qubit gates, respectively. Here, $p_{i_k}$ is the corresponding error probability, $U_i$ represents a layer of certain gates, and $\rho$ is the original $n$-qubit state.
\end{definition}

Operating under this layer-by-layer noise model, we can arrive at our second theorem, which provides an exact expression for the error.
\begin{theorem}
    Let a quantum circuit consist of $n$ qubits, with $d_1$ layers of single-qubit gates and $d_2$ layers of two-qubit gates. Assume the presence of a global depolarizing noise in the circuit, with error probabilities $p_1$ for single-qubit gates and $p_2$ for two-qubit gates. Let $\rho$ denote the quantum state corresponding to the ideal circuit. Then, the error $\Delta(\rho)$ between the ideal and noisy measurements satisfies
    \begin{equation}
        \begin{aligned}
            \Delta(\rho) &= \left[1-(1-p_{1})^{d_1}(1-p_2)^{d_2}\right]\\
            &\quad\times\sqrt{\sum_{p q r s}\left[\text{Tr}(\rho a_p^\dagger a_q^\dagger a_r a_s)-\frac{1}{2^n}\text{Tr}(a_p^\dagger a_q^\dagger a_r a_s)\right]^2},
        \end{aligned}
    \end{equation}
    where $a_i^\dagger$ and $a_i$ represent the creation and annihilation operators for the $i$-th qubit, respectively.
    \label{thm:2}
\end{theorem}

The proof of Theorem~\ref{thm:1} is as follows.
\begin{proof}
    By definition, we can write the ideal single-qubit gate as
    \begin{equation}
        \Omega_i^k\left(\rho_0\right)=U_i \rho_0 U_i^{\dagger},
    \end{equation}
    representing the $i$th gate on $k$th qubit.
    Then, the depolarizing noise affected single-qubit gate channel can be written as
    \begin{equation}
        \Lambda_i^k\left(\rho_0\right)=\left(1-p_{k_1}\right) U_i \rho_0 U_i^{\dagger}+p_{k_1}/2 \operatorname{Tr}_k\left(\rho_0\right) \otimes I
    \end{equation}
    The two-qubit gate situation is similar.

    \begin{equation}
        \Delta(\rho)=\sqrt{\sum_{pqrs}\left(\operatorname{Tr}\left(\rho^{\prime} a_p^{\dagger} a_q^{\dagger} a_r a_s\right)-\operatorname{Tr}\left(\rho_1 a_p^{\dagger} a_q^{\dagger} a_r a_s\right)\right)^2}
        \label{eq:a3}
    \end{equation}
    where we define
    \begin{equation}
        \rho^{\prime}=\Lambda_N \cdots \Lambda_1\left(\rho_0\right)
    \end{equation}
    and
    \begin{equation}
        \rho_k=\Omega_N \cdots\Omega_k\Lambda_{k-1} \cdots \Lambda_1\left(\rho_0\right).
    \end{equation}
    Since
    \begin{equation}
        \begin{aligned}
            & \operatorname{Tr}\left[\Lambda(\rho) a_p^{\dagger} a_q^{\dagger} a_r a_s\right]-\operatorname{Tr}\left[\Omega(\rho) a_p^{\dagger} a_q^{\dagger} a_r a_s\right] \\
            = & \operatorname{Tr}\left\{p_k\left[\operatorname{Tr}_k(\rho) \otimes I / 2-U_i \rho U_i^{\dagger}\right] a_p^{\dagger} a_q^{\dagger} a_r a_s\right\} \\
            \leqslant & 2 p_k, \quad k=1,2,
\end{aligned}
    \end{equation}
    we can bound the summation term in Eq.~\ref{eq:a3} to be
    \begin{equation}
        \begin{aligned}
            & \operatorname{Tr}\left(\rho^{\prime} a_p^{\dagger} a_q^{\dagger} a_r a_s\right)-\operatorname{Tr}\left(\rho_1 a_p^{\dagger} a_q^{\dagger} a_r a_s\right) \\
            = & \operatorname{Tr}\left(\rho^{\prime} a_p^{\dagger} a_q^{\dagger} a_r a_s\right)-\operatorname{tr}\left(\rho_N a_p^{\dagger} a_q^{\dagger} a_r a_s\right)+\cdots \\
            + & \operatorname{tr}\left(\rho_2 a_p^{\dagger} a_q^{\dagger} a_r a_s\right)-\operatorname{tr}\left(\rho_1 a_p^{\dagger} a_q^{\dagger} a_r a_s\right)\\
            \leqslant & 2\sum_{k_1,k_2}^{n_1,n_2}(p_{k_1}+p_{k_2}).
        \end{aligned}
    \end{equation}
    Thus, 
    \begin{equation}
        \Delta(\rho)\leqslant 2n^2\sum_{k_1,k_2}^{n_1,n_2}(p_{k_1}+p_{k_2})
    \end{equation}
\end{proof}

The proof of Theorem~\ref{thm:2} is as follows.
\begin{proof}
    By definition, one layer of single-qubit gates of weakened depolarizing noise channel can be written as
    \begin{equation}
        \rho_1=\Phi_1\left(\rho_0^{\prime}\right)=\left(1-p_{k_1}\right) U_1 \rho_0 U_1^{\dagger}+p_{k_1} \frac{I_n}{2^n}.
    \end{equation}
    The effect of imposing a second layer of channels on top of this is given by
    \begin{equation}
        \begin{aligned}
            \rho_2 = & \Phi_2\left(\rho_1\right)=\left(1-p_{k_2}\right) \rho_1+p_{k_2} \frac{I_n}{2^n} \\
            = & \left(1-p_{k_1}\right)\left(1-p_{k_2}\right) U_2 U_1 \rho_0 U_1^{\dagger} U_2^{\dagger}\\
            & +\left[\left(1-p_{k_2}\right) p_{k_1}+p_{k_2}\right] \frac{I_n}{2^n}.
\end{aligned}
    \end{equation}
    After all it becomes
    \begin{equation}
    \begin{aligned}
        \rho^\prime= & \left(1-p_1\right)^{d_1}\left(1-p_2\right)^{d_2} U \rho_0 U^{\dagger}\\
        & +\left[1-\left(1-p_1\right)^{d_1}\left(1-p_2\right)^{d_2}\right] \frac{I_n}{2^n},
    \end{aligned}
    \end{equation}
    where $U=\prod_i U_i$.
    The ideal case is
    \begin{equation}
        \rho = U\rho_0 U^\dagger.
    \end{equation}
    Eventually, by definition of the error, we have
    \begin{equation}
        \begin{aligned}
            \Delta(\rho) &= \left[1-(1-p_1)^{d_1}(1-p_2)^{d_2}\right]\\
            &\quad\times\sqrt{\sum_{p q r s}\left[\Tr(\rho a_p^\dagger a_q^\dagger a_r a_s)-\frac{1}{2^n}\Tr(a_p^\dagger a_q^\dagger a_r a_s)\right]^2}.
        \end{aligned}
    \end{equation}
\end{proof}

\end{widetext}

\end{document}